\documentclass[prd,twocolumn,amssymb,10pt]{revtex4}
\usepackage{amsmath,amsfonts,amssymb}
\usepackage{verbatim,graphicx,float}  

\begin{document}

\title{Multi-fluid potential in the loop cosmology}

\author{Jakub Mielczarek}
\email{jakub.f.mielczarek@gmail.com}
\affiliation{Astronomical Observatory, Jagiellonian University, 30-244
Krak\'ow, Orla 171, Poland}


\begin{abstract}
The scalar field can behave like a fluid with equation of state $p_{\phi}=w\rho_{\phi}$, where 
$w \in [-1,1]$. In this Letter we derive a class of the scalar field potentials for which 
$w=$ const. Scalar field with such a potential can mimic ordinary matter, radiation, cosmic strings, etc.   
We perform our calculations in the framework of the loop cosmology
with holonomy corrections. We solve the model analytically for the whole parameter space. 
Subsequently, we perform similar consideration for the model with a phantom field ($w<-1$).
We show that scalar field is monotonic function in both cases. This indicates that it can 
be treated as a well-defined internal time for these models.  Moreover we perform 
preliminary studies of the scalar field perturbations with this potential. We indicate that 
non-Gaussian features are present admitting for the possible observational constraints of the model. 
\end{abstract}

\maketitle

\section{Introduction} \label{sec:intro}

In the framework of Loop Quantum Cosmology (LQC)\cite{Bojowald} one can introduce phenomenological 
Hamiltonian \cite{Ashtekar:2006rx,Ashtekar:2006uz,Ashtekar:2006wn}
\begin{eqnarray}
\mathcal{H}_{\text{phen}} &=&  - \frac{3}{8 \pi G \gamma^2} \sqrt{|p|} \left[ \frac{ \sin \left( \bar{\mu} c
\right) }{\bar{\mu}}   \right]^2  \nonumber \\
&+& \frac{1}{2} \frac{ \pi_{\phi}^2}{  {|p|}^{3/2} } +{|p|}^{3/2}V(\phi).
\label{model}
\end{eqnarray}
Here effects of the quantum holonomies have been introduced. The classical limit corresponds to 
the case $\bar{\mu}\rightarrow 0$, what gives $\sin \left( \bar{\mu} c \right)/\bar{\mu} \rightarrow c$.  
The parameter $\bar{\mu}$ is in general function of the canonical variable $p$. In our considerations 
we choose $\bar{\mu}=\sqrt{\Delta/|p|}$ where $\Delta=2\sqrt{3}\pi\gamma l^2_{\text{Pl}}$. 
It was shown that this leads to the proper classical limit and can be treated as 
a unique choice in the loop quantization class \cite{Corichi:2008zb}.
Based on the Hamilton equation $\dot{f} = \{f,\mathcal{H}_{\text{phen}}\}$ we can derive 
equations of motion for the canonical 
variables $(c,p,\phi,\pi_{\phi})$. The variables $(c,p)$ can be related to the standard FRW 
variables $(c,|p|)=(\gamma \dot{a} V^{1/3}_0,a^2V^{2/3}_0 )$. Subsequently, with use of the 
Hamiltonian constraint $\mathcal{H}_{\text{phen}}=0$ we can derive modified Friedmann equation 
\begin{equation}
H^2\equiv \left( \frac{\dot{p}}{2p}\right)^2 = 
\frac{\kappa}{3} \rho_{\phi} \left(1 -\frac{\rho_{\phi}}{\rho_{\text{c}}} \right)
\label{Friedmann}
\end{equation}
where $\kappa=8\pi G$ and
\begin{equation}
\rho_{\text{c}} =  \frac{\sqrt{3}}{16\pi^2 \gamma^3 l_{\text{Pl}}^4}.
\end{equation} 
The equation of motion of the scalar field $\phi$ holds classical form
\begin{equation}
\ddot{\phi}+3H\dot{\phi}+\frac{dV}{d\phi}= 0.
\label{fieldequation}
\end{equation}

The special limit of the above theory is the case $V(\phi)=0$, then 
$\mathcal{H}_{\text{phen}}\rightarrow \mathcal{H}_{\text{eff}}$.  It means that 
solution of the equation (\ref{Friedmann}) traces the mean value $\langle \hat{p} \rangle$ 
(for rigorous verification of the effective scenario in LQC see Ref. \cite{Taveras:2008ke}). 
It is however not necessarily the true for the general case with $V(\phi) \neq 0$.
However, even when we are not sure whether theory is truly effective one, it is
interesting to investigate phenomenological consequences of the models with 
potentials. Such studies have been performed for the broad class of potentials.
In Ref. \cite{Mielczarek:2008zv} model (\ref{model}) with constant potential
($V=\Lambda/8\pi G$) has been solved analytically. Power law, cyclic and bicyclic potentials have been 
numerically studied in Ref. \cite{Singh:2006im}. Recently model with exponential potential has 
been studied, exhibiting sudden singularity behaviour \cite{Cailleteau:2008wu}.
Other examples can be found in Ref. \cite{Mielczarek:2008ps,Artymowski:2008sc}.
     
In this Letter we perform restriction for the scalar field energy density and pressure 
$p_{\phi}=w\rho_{\phi}$ with $w =$ const. In this situation scalar field can mimic 
any matter with equation of state $p=w\rho$ where $w =$ const and $w \in [-1,1]$.
Based on this restriction we derive expression for the scalar field 
potential which fulfills this condition. We repeat these considerations for the 
case of phantom field $(w<-1)$.

Subsequently we highlight possible applications of the model. In particular we consider 
scalar field perturbations and indicate its non-Gaussian features. This can be 
important from the point of observational constraints for the model. Namely 
since the obtained potential is non-quadratic, interaction between different modes 
is present, leading to the nonlinear effects in the CMB radiation. 
Therefore presented model can be potentially verified with astronomical observations.

We show that scalar field with derived potential is monotonic function of coordinate time.
This is important property since in quantum cosmological models we are looking for a well-defined 
intrinsic time. This is necessary since the Hamiltonian constraint $\mathcal{H}=0$ 
leads to the time coordinate independent Wheeler-DeWitt equation $\hat{\mathcal{H}}|\Psi\rangle=0$. 
Therefore wave function of the universe $|\Psi\rangle$ does not depend on time explicitly
and some new  intrinsic time must be introduced to trace quantum evolution.
Monotonic scalar field is good candidate. Namely performing 
canonical quantisation we replace $\pi_{\phi}\rightarrow \hat{\pi}_{\phi}=-i\partial_{\phi}$. 
Then matter part of the Hamiltonian is 
\begin{equation}
\hat{\mathcal{H}}_{\text{m}}=-\frac{1}{2{|p|}^{3/2} }\frac{\partial^2}{\partial \phi^2} +{|p|}^{3/2}V(\phi).
\end{equation}      
Now $\phi$ plays a role of time in analogy with the Klein-Gordon equation.

\section{Multi-fluid potential} \label{sec:Multi}

\begin{figure}[ht!]
\centering
\includegraphics[width=7cm,angle=0]{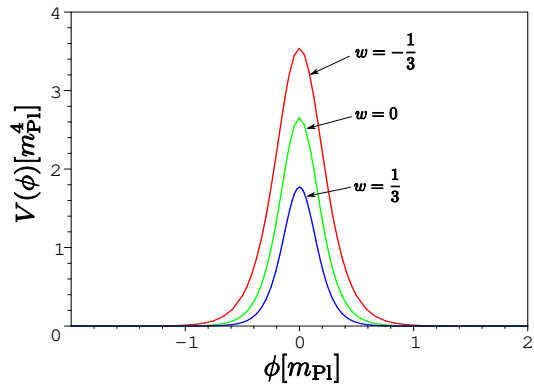}
\caption{Multi-fluid potential with the different values of the parameter $w$. In the figure
the value $\phi_0=0$ was assumed.}
\label{Pot}
\end{figure}

Energy density and pressure of the homogeneous scalar field are expressed as follows 
\begin{eqnarray}
\rho_{\phi} &=& \frac{1}{2}\dot{\phi}^2+V(\phi), \label{defrhophi}  \\
p_{\phi} &=& \frac{1}{2}\dot{\phi}^2-V(\phi).
\end{eqnarray}
In this section we consider restriction $p_{\phi}=w\rho_{\phi}$ where $w=$ const.
This assumption leads to the expression $\frac{1}{2}\dot{\phi}^2=\frac{1+w}{1-w}V(\phi)$ which gives
\begin{eqnarray}
\rho_{\phi} &=& \frac{2}{1-w} V(\phi), \label{rhoV}  \\
p_{\phi} &=& \frac{2w}{1-w}V(\phi) .
\end{eqnarray}
With use of equations (\ref{Friedmann}),(\ref{fieldequation}) and considered condition we obtain 
\begin{equation}
\left(\frac{dV(\phi)}{d\phi}\right)^2 =3\kappa(1+w)V(\phi)^2\left( 1- \frac{2}{1-w}\frac{V(\phi)}{\rho_c}\right). 
\end{equation}
The above equation has solution in the form
\begin{equation}
V(\phi) =  \frac{V_*}{\cosh^2\left[\sqrt{6\pi G (1+w)}(\phi-\phi_0)\right] } 
\label{Potential}
\end{equation}
where $V_*=\rho_c (1-w)/2$. It is worth to mention that for $\phi \rightarrow \infty$ we recover 
$V(\phi) \propto \exp \{ -\sqrt{24\pi G (1+w)}\phi \}$ which is well-known classical limit.
In Fig. \ref{Pot} we draw potential (\ref{Potential}) for the different values of parameter $w$.

Particular situation corresponds to the case $p_{\phi}=-\rho_{\phi}$ $(w=-1)$. 
Field can get such an equation of state only in the limit $\dot{\phi} \rightarrow 0$.
Therefore this particular case  must be excluded from considerations. Namely, we 
cannot mimic pure cosmological constant with time dependent scalar field. 

We must to stress that potential in the form $1/\cosh^2(x)$ has already appeared 
in the different context in the loop cosmology. Namely, it was obtained by Singh 
\cite{Singh:2006sg} for the scaling solutions in LQC
dual to those in Randall-Sundrum cosmology.  These considerations were performed for the 
model with a self-interacting scalar field and matter with a fixed equation of state $w$. 

\section{Analytical solutions} \label{sec:analytical}

In this section we show analytical solution of the model with potential (\ref{Potential}).
For the considered case we can additionally derive equation
\begin{equation}
\dot{H}=-\frac{\kappa}{2}(1+w)\rho_{\phi}\left(1-2\frac{\rho_{\phi}}{\rho_c} \right).
\end{equation}
With use of continuity equation $\dot{\rho}_{\phi}+3H(\rho_{\phi}+p_{\phi})=0$ 
(or equivalently equation (\ref{fieldequation}) ) we directly obtain
\begin{equation}
\rho_{\phi} = \rho_c \left(\frac{p}{p_c}\right)^{-\frac{3}{2}(1+w)}.
\end{equation}
Comparing this to the expression (\ref{rhoV}) with (\ref{Potential}) we obtain
\begin{equation}
p(\phi) = p_c \left( \cosh^2\left[\sqrt{6\pi G (1+w)}(\phi-\phi_0)\right]   \right)^{\frac{2}{3(1+w)}}. \label{Solp}
\end{equation}
Similar solution, however not in such an explicit form, has already been derived in Ref. \cite{Singh:2006sg}.
To obtain time $t$ dependence we use  expression $\frac{1}{2}\dot{\phi}^2=\frac{1+w}{1-w}V(\phi)$
with potential (\ref{Potential}) to obtain 
\begin{equation}
\cosh^2\left[\sqrt{6\pi G (1+w)}(\phi-\phi_0)\right] = 1+6\pi G \rho_c (1+w)^2 (t-t_0)^2. \label{phi_t}
\end{equation}
This leads to
\begin{equation}
p(t) = p_c \left( 1+6\pi G \rho_c (1+w)^2 (t-t_0)^2 \right)^{\frac{2}{3(1+w)}}. \label{solpt}
\end{equation}
This solution can be also found in Ref. \cite{Artymowski:2008sc}.
In case of the $w=-1$ solution has exponential form and can be found in the paper \cite{Mielczarek:2008zv}.

\begin{figure}[ht!]
\centering
\includegraphics[width=7cm,angle=0]{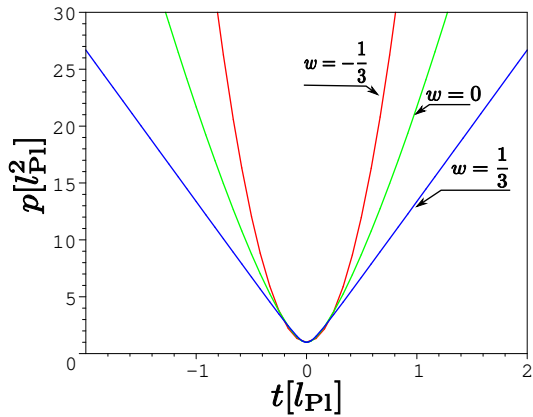}
\caption{Evolution of the canonical variable $p$ for the different values of parameter $w$.}
\label{bounce}
\end{figure}

In Fig. \ref{bounce} we show time dependence of the canonical variable $p$ for different 
values of parameter $w$.
Resulting dynamics is singularity free bounce.

\begin{figure}[ht!]
\centering
\includegraphics[width=7cm,angle=0]{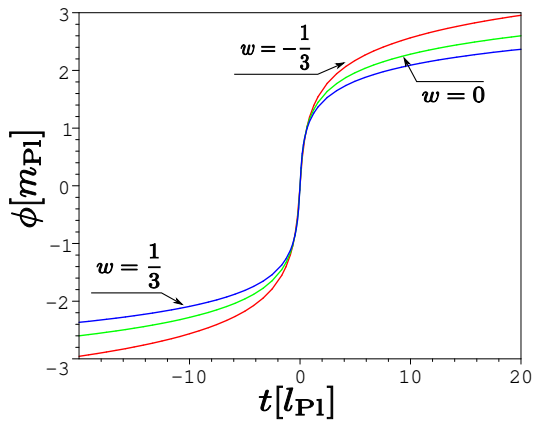}
\caption{Evolution of the scalar field $\phi$ for the different values of parameter $w$.}
\label{field}
\end{figure}

The time derivative of the field $\phi$ can be obtained from the equation (\ref{phi_t}) and has a form
\begin{equation}
\frac{d\phi}{dt} =\pm \sqrt{\frac{\rho_c(1+w)}{1+6\pi G \rho_c (1+w)^2(t-t_0)^2}}. 
\end{equation}
From this we see that $\phi$ is monotonic function of time (growing or decreasing).
Choosing growing solution we have
\begin{equation}
\phi(t)=\phi_0+\frac{ \text{arcsinh} \left[ \sqrt{6\pi G \rho_c}(1+w) (t-t_0) \right]}{\sqrt{6\pi G(1+w)}}. \label{solphit}
\end{equation}
We draw this function in Fig. \ref{field}.

To complete above reasoning we verify obtained solutions in Appendix \ref{sec:appendix}.
Namely we show there how to solve equations of motion without imposing the condition $w =$ const in 
equation of state. Instead of this we solve equations from scratch assuming the form of potential (\ref{Potential}).

\section{Phantom field} \label{sec:phantom}

In this section we perform analogous consideration for the case of phantom field \cite{Caldwell:1999ew,Singh:2003vx,
Dabrowski:2003jm}. For this kind of field energy density and pressure are expressed as follows
\begin{eqnarray}
\rho_{\phi} &=& -\frac{1}{2}\dot{\phi}^2+V(\phi),  \\
p_{\phi} &=& -\frac{1}{2}\dot{\phi}^2 - V(\phi).
\end{eqnarray}
From this we see that $\rho_{\phi}+p_{\phi}< -\dot{\phi}^2$. The parameter $w$ in the equation of
state for the phantom field takes values $w<-1$. 

Phantom matter was introduced in loop quantum cosmology in Ref. \cite{Sami:2006wj}.
Subsequently phantom field in the loop quantum cosmology has been studied in Ref. \cite{Samart:2007xz}.
Also the models with interacting phantom field have been investigated \cite{Gumjudpai:2007fc,Fu:2008gh,Wu:2008db}.
 
Based on the continuity equation we derive equation of motion
\begin{equation}
\ddot{\phi}+3H\dot{\phi}-\frac{dV}{d\phi}= 0.
\label{fieldequation2}
\end{equation}
Performing calculations in the same way like in the section  \ref{sec:Multi} we obtain 
\begin{equation}
V(\phi) =  \frac{V_*}{\cosh^2\left[\sqrt{-6\pi G (1+w)}(\phi-\phi_0)\right] } 
\label{Potential2}
\end{equation}
where $V_*=\rho_c (1-w)/2$.

\begin{figure}[ht!]
\centering
\includegraphics[width=7cm,angle=0]{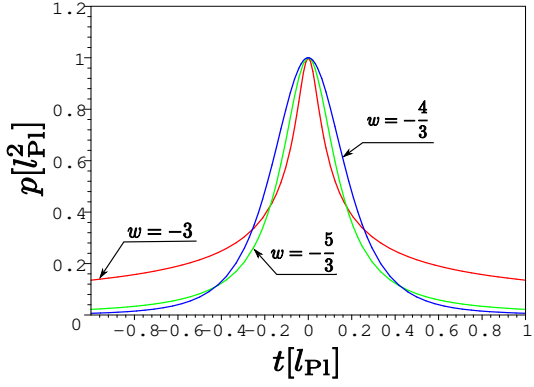}
\caption{Evolution of the canonical variable $p$ for the different values of parameter $w$.}
\label{phan}
\end{figure}

\begin{figure}[ht!]
\centering
\includegraphics[width=7cm,angle=0]{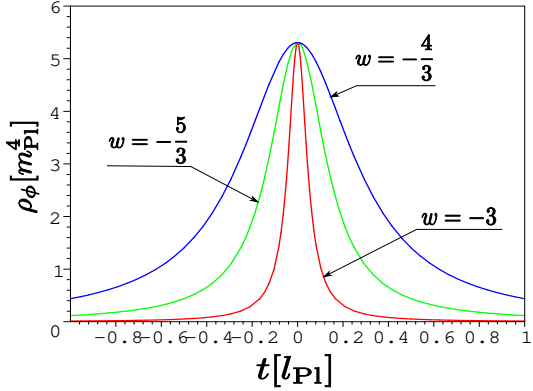}
\caption{Evolution of the energy density of the phantom field $\rho_{\phi}$ 
for the different values of parameter $w$.}
\label{rho}
\end{figure}

Solution for the scalar phantom field has a form
\begin{equation}
\phi(t)=\phi_0+\frac{ \text{arcsinh} \left[ \sqrt{6\pi G \rho_c}|1+w| (t-t_0) \right]}{\sqrt{-6\pi G(1+w)}}. 
\label{phiphantom}
\end{equation}
It is monotonically growing function too. Therefore phantom field with potential (\ref{Potential2})
can be treated as internal time. The canonical parameter $p$ depends on value of the $\phi$ as follows 
\begin{equation}
p(\phi) = p_c \left( \cosh^2\left[\sqrt{-6\pi G (1+w)}(\phi-\phi_0)\right]   \right)^{\frac{2}{3(1+w)}}.
\end{equation}
Rewriting equation (\ref{phiphantom}) to the form 
\begin{equation}
\cosh^2\left[\sqrt{-6\pi G (1+w)}(\phi-\phi_0)\right] = 1+6\pi G \rho_c (1+w)^2 (t-t_0)^2
\end{equation}
one can obtain explicit time dependence of the parameter $p$. We show this function 
in Fig. \ref{phan}. 
We see that, in contrast to the bouncing behaviour, here universe shrinks to zero for $t \rightarrow \pm \infty$. 
To investigate this state we consider time dependence of the energy density of the phantom field.  
We have 
\begin{equation}
\rho_{\phi} = \frac{\rho_{c}}{1+6\pi G \rho_c (1+w)^2 (t-t_0)^2}
\end{equation}
what is shown in Fig. \ref{rho}.
We see that despite the variable $p$ tends to zero energy density decreases too.

\section{Perturbations and non-Gaussianity} \label{sec:perturbations}

In the previous part of this Letter we have considered models with a homogeneous scalar field.
Now we are going to take into account perturbations. We consider simplified model where 
scalar field is perturbed while background geometry holds its homogeneity. 
Moreover we aim rather to indicate further applications of the potential (\ref{Potential})
than to perform full analysis. Therefore considerations presented should be 
seen as an invitation to the more detailed studies.

We split the scalar field $\phi({\bf x },t)$ as 
\begin{equation}
\phi({\bf x },t) = \bar{\phi}(t)+\delta\phi({\bf x },t).
\end{equation}
Here $\bar{\phi}(t)$ is homogeneous mode defined as 
\begin{equation}
\bar{\phi}(t)=\frac{1}{V_0}\int_{V_0}d^3{\bf x} \phi({\bf x },t)
\end{equation}
where $V_0$ is some fixed fiducial volume.
Then equation for the background part is
\begin{equation}
\ddot{\bar{\phi}}+3H\dot{\bar{\phi}}+\frac{dV(\bar{\phi})}{d\bar{\phi}} = 0.
\end{equation}
Therefore solutions of equations of motion for the background part are those 
derived in Section \ref{sec:analytical} with $\phi=\bar{\phi}$. 
First order perturbation of the full 
equation of motion leads to 
\begin{equation}
\ddot{\delta\phi}+3H\dot{\delta\phi}-\frac{1}{a^2}\nabla^2 \delta\phi +m^2(t)\delta\phi  = 0
\end{equation}
which is valid for $|\delta \phi/\phi|\ll 1$.
Here we have defined the mass term
\begin{equation}
m^2(t) = \left.\frac{d^2 V}{d\phi^2} \right|_{\phi=\bar{\phi}}= 6\pi G \rho_c (1-w^2) \frac{2B(t)-3}{B^2(t)}
\end{equation}
where 
\begin{equation}
B(t) \equiv \cosh^2\left[\sqrt{6\pi G (1+w)}\bar{\phi}\right] = 1+6\pi G \rho_c (1+w)^2 t^2. \label{phi_t}
\end{equation}
Performing Fourier transform 
\begin{equation}
\delta \phi({\bf x },t) = \int \frac{d^3{\bf k}}{(2\pi)^{3/2}}  \frac{u({\bf k },t)}{a(t)} e^{i {\bf k \cdot x}}
\end{equation}
and introducing conformal time $d\eta = dt/a(t)$ we can write equation for the 
mode functions $u({\bf k },\eta)$ as
\begin{equation}
\frac{d^2}{d\eta^2}u({\bf k },\eta)+\left[k^2+m^2_{\text{eff}}  \right]u({\bf k },\eta)=0.  \label{equ} 
\end{equation}
Here we have defined effective mass 
\begin{equation}
m^2_{\text{eff}}=a^2m^2-\frac{a^{''}}{a}  =  a^2\left[m^2-H^2-\frac{\ddot{a}}{a}\right]. \label{meffDef}
\end{equation}
Based on the equation (\ref{Solp}) we can write solution for the scale factor as follows 
\begin{equation}
a(t) = a_c \left( 1+6\pi G \rho_c (1+w)^2 t^2 \right)^{\frac{1}{3(1+w)}}.
\end{equation}
Applying this to the equation (\ref{meffDef}) we obtain
\begin{eqnarray}
m^2_{\text{eff}} &=& a_c^2\frac{1}{8}\kappa^2\rho_c^2(1+w)^2(8+3w-9w^2)B(t)^{\alpha} \times \nonumber \\
                &\times& \left[t^2-\frac{2}{3}\frac{\Delta\gamma^2(5+2w-3w^2)}{(1+w)^2(8+3w-9w^2)}\right] 
\label{meff2}
\end{eqnarray}
where 
\begin{equation}
\alpha =-\frac{2}{3} \frac{2+3w}{1+w}. 
\end{equation}
The full analysis of the equation (\ref{equ}) with $m^2_{\text{eff}}$ given above is 
behind the scope of this Letter. However we can immediately investigate its behaviour in the 
pre-bounce phase $t\rightarrow -\infty$. In this limit
\begin{equation}
m^2_{\text{eff}} \rightarrow  \frac{16+6w-18w^2}{(1+3w)^2} \frac{1}{\eta^2}
\end{equation}
where we have changed time for conformal. Advanced and normalised solution of the equation (\ref{equ}) in
 considered limit is
\begin{equation}
u(k,\eta) = \sqrt{-k\eta} \sqrt{\frac{\pi}{4k}} e^{i\frac{\pi}{2}\left(|\nu|+\frac{1}{2}\right)} H^{(1)}_{|\nu|}(-\eta k) 
\end{equation}
where
\begin{equation}
\nu^2=\frac{9}{4}\frac{(9w^2-2w-7)}{(1+3w)^2}.
\end{equation}
This solution is valid for $\nu^2\geq 0$ or equivalently  for $w \in [-1,-7/9] \cup \{1\}$.
Therefore only in the small range. Power spectrum of perturbations is then given as 
\begin{equation}
\mathcal{P}_{\delta\phi} = \frac{k^3}{2\pi^2} \left| \frac{u}{a} \right|^2 \propto (-k\eta)^{3-2|\nu|}
\end{equation}
where super-horizontal approximation
\begin{equation}
H^{(1)}_{n}(x) \simeq -\frac{i}{\pi} \Gamma(n) \left(\frac{x}{2}\right)^{-n} \ \ \text{for} \ \ x \ll 1
\end{equation}
has been used. It is worth to mention that scale invariant spectrum $|\nu|=\frac{3}{2}$ is recovered only 
for $w=-1$. Therefore duality investigated in Ref. \cite{Wands:1998yp} is broken here due 
to additional mass term.  In the range
$w \in (-7/9,1)$ ($\nu^2 < 0$) we can obtain solutions by the replacement $\nu=-i\tilde{\nu}$. 
Then for $-k\eta \ll 1$ we have $\mathcal{P}_{\delta\phi} \propto k^3$ as discussed in 
Ref. \cite{Riotto:2002yw}. 

Presented solutions are valid in the pre-bounce phase. Later when modes 
cross the bounce solutions of the equation of motion have a different form. 
Despite the fact that size of horizon diverges at the turning point 
($H\rightarrow 0$ $\Rightarrow$ $1/|H|\rightarrow \infty$) 
the solutions evolve smoothly without any anomalies. In the context of loop cosmology such a 
behaviour has been recently approved for gravitational waves based on the analytical and numerical 
calculations \cite{Mielczarek:2009vi}. Interesting generic feature of 
the bouncing cosmologies is suppression of power of the low energy models 
\cite{Piao:2003zm,Mielczarek:2008pf}. 
Therefore we can expect damping of the low energy modes also for the model studied here. 
However it can depend whether condition $|\delta \phi/\phi|\ll 1$ for validity of
the linear approximation  is conserved during the whole evolution.
If this condition is not satisfied then nonlinear effects should be also 
taken into account. Otherwise suppression of power of the low energy modes
should be easily obtained. Such an effect is typically presented as a possible 
explanation of the low CMB multipoles suppression \cite{Contaldi:2003zv}. 
However it seems that fine tuning must occur to match energy scales of the bounce 
with the cosmological ones at which suppression of the CMB modes is observed. These issues have to be
addressed in the further considerations. 

Scalar field with potential (\ref{Potential}) is non-Gaussian. Namely modes with different 
wave numbers interact one another \cite{Bartolo:2004if}. Since observations suggest that primordial perturbations 
were nearly Gaussian \cite{Komatsu:2008hk} the CMB nonlinearities can in principle be used to restrict 
models with interacting potentials. In our case potential of interaction has
non-polynomial form therefore non-Gaussian features of the field should be significant.
However since we do not know how many perturbations originate from the 
short quantum phase (where model studied here can be applied) we cannot infer about 
their impact on the CMB nonlinearities. In principle contribution from the phase of bounce can be 
affected by the succeed phase of inflation. Detailed studies of this issues are however beyond 
the scope of this Letter. We devote for these issues another paper \cite{MielczarekX2009} 
while now concentrate only on some preliminary estimations of nonlinear effects.
    
\begin{figure}[ht!]
\centering
\includegraphics[width=7cm,angle=0]{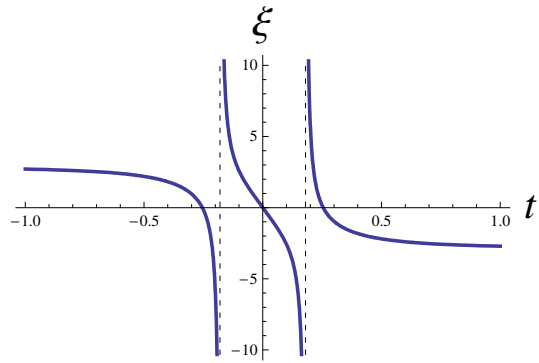}
\caption{Parameter of non-Gaussianity $\xi$ for $w=0$.}
\label{Xi}
\end{figure}

The lowest order interacting term is given by 
\begin{equation}
V_{\text{int}} = \frac{\Lambda(t)}{3!}(\delta \phi)^3+\mathcal{O}\left((\delta \phi)^4\right)
\end{equation}
where 
\begin{eqnarray}
\Lambda(t) &=& \left.\frac{d^3 V}{d\phi^3} \right|_{\phi=\bar{\phi}}= 24 \ \text{sgn}(t) \sqrt{6} \pi ^{3/2}\rho_c(1-w)
 \times \nonumber \\  &\times&  ((1+w)G)^{3/2}\frac{3-B(t)}{B^2(t)}\sqrt{\frac{B(t)-1}{B(t)}}.
\end{eqnarray}
This term becomes important when dominates the mass one. Therefore ratio  
\begin{equation}
\frac{\frac{\Lambda(t)}{3!}(\delta \phi)^3}{\frac{m^2(t)}{2!}(\delta \phi)^2} = 
\frac{1}{3} \frac{\Lambda(t)}{m^2(t)} \delta \phi = \xi(t) \delta \phi
\end{equation}
can be seen as a measure of non-Gaussianity. Here we have defined parameter 
\begin{equation}
\xi(t)= \text{sgn}(t) \sqrt{\frac{4}{3}\kappa(1+w)} \frac{3-B(t)}{2B(t)-3}\sqrt{\frac{B(t)-1}{B(t)}}.
\end{equation}
We show this function in Fig. \ref{Xi}.
At the points where $B(t)=\frac{3}{2}$ non-Gaussian part becomes dominant since the mass vanishes there.   
In the limits
\begin{equation}
\lim_{t\rightarrow\pm \infty}\xi(t)= \mp \sqrt{\frac{1}{3}\kappa(1+w)}
\end{equation}
the parameter $\xi$ stabilises.
Above results indicate that non-Gaussian effects can be in fact produced. Especially interesting 
are points where $B(t)=\frac{3}{2}$, then nonlinear effects are dominant. 
These regions could be an effective source of the cosmological non-Gaussianity.
It is therefore in principle possible to constraint this model from CMB nonlinearities.
This could bring observational bounds on parameter $\rho_c$ or equivalently 
Barbero-Immirzi parameter $\gamma$ in this specific model. However to prove 
this supposition further detailed analysis is required. In principle 
bispectrum and resulting parameter of nonlinearities $f_{NL}$ have to be 
determined.
 
\section{Summary} \label{sec:summary}

In this Letter we have considered self-interacting scalar field theory in the universe 
with holonomy effects resulting from the Loop Quantum Cosmology. We have performed constraint for the
scalar field in the form $p_{\phi}=w\rho_{\phi}$ where $w=$const.
Based on this restriction we have derived expression for the scalar field 
potential which fulfills this condition.  This condition can be fulfilled for
$w\in (-1,1]$. The case $w=-1$ is excluded. Such a state can be obtained only in the 
case $\dot{\phi} \rightarrow 0$ and cannot be mimicked with $\dot{\phi} \neq 0$.
We have solved considered model analytically. Subsequently, we have performed analogous 
investigations for the phantom field $w <-1$. Potentials for scalar and phantom fields are 
related by the symmetry $(1+w) \rightarrow -(1+w)$.

We have shown how barotropic matter can be introduced in LQC in the well-defined manner. 
In this approach matter content is mimicked by the scalar field with derived potential (\ref{Potential}).
We have shown that field $\phi$ is monotonic function of time in both considered cases. This result 
indicates that scalar field (as well as phantom field) with derived potentials can
be moreover treated as intrinsic time. The models therefore seems to be good candidate 
to the further purely quantum considerations.  

We have performed preliminary studies of the perturbations of the scalar field with 
obtained multi-fluid potential. In the pre-bounce phase super-horizontal modes 
are typically damped. Namely the power spectrum is $\mathcal{P}_{\delta\phi} \propto k^3$.
However to verify that this spectrum survives during the bounce phase further numerical 
analysis should be performed. If super-horizontal spectrum does not change qualitatively this can 
lead to the suppression of the low CMB multipoles. We have also indicated that non-Gaussian 
perturbations are produced. Further analysis of this issue may provide observational 
bounds on the model. Therefore beside the purely theoretical results as the method of introducing 
barotropic mater in LQC also phenomenological consequences were found. 

\begin{acknowledgments}
Author would like to thank to anonymous referee for 
useful remarks.
This work was supported in part by the Marie Curie Actions Transfer of
Knowledge project COCOS (contract MTKD-CT-2004-517186).
\end{acknowledgments}

\appendix
\section{Solution  of the equations of motion - verification} \label{sec:appendix}

In this appendix we are going to  verify obtained solutions (\ref{solpt},\ref{solphit}) solving 
from scratch equations of motion (\ref{Friedmann},\ref{fieldequation}) with potential (\ref{Potential}).  
Without loose of generality we assume here $\phi_0=0$. Therefore we consider a theory 
with potential in the form
\begin{equation}
V(\phi) =  \frac{V_*}{\cosh^2\left[\sqrt{6\pi G (1+w')}\phi\right]} 
\label{AP0}
\end{equation}
where $V_*=\rho_c (1-w')/2$. Here we do not assume what is the interpretation of the parameter $w'$. 
It is just some fixed parameter and its meaning will be clarified later. 
Equation (\ref{fieldequation}) with potential (\ref{AP0}) takes the form
\begin{equation}
\ddot{\phi}+3H\dot{\phi}-2V_*\alpha \frac{\tanh(\alpha \phi) }{\cosh^2(\alpha\phi )} = 0 \label{AP1}
\end{equation}
where to simplify notation we have defined $\alpha=\sqrt{6\pi G (1+w')}$.
To solve this equation we postulate solution in the form
\begin{equation}
\phi(t)=\frac{1}{\alpha} \text{arcsinh} \left(\beta t \right) \label{AP2}
\end{equation}
where $\beta$ is some unknown constant. Applying this solution, we rewrite 
equation (\ref{AP1}) to the form 
\begin{equation}
 H = \frac{t}{1+\beta^2t^2} \left[2 V_*\alpha^2+\beta^2 \right]\frac{1}{3}.  \label{AP3}
\end{equation}
Now let us calculate expression for the energy density (\ref{defrhophi}). With use 
of (\ref{AP2}) we obtain
\begin{equation}
\rho =   \frac{t}{1+\beta^2t^2} \left[ \frac{\beta^2}{2\alpha^2}+V_* \right]. 
\end{equation}
Applying this to the modified Friedmann equation (\ref{Friedmann}) we have
\begin{equation}
H^2=\frac{\kappa}{3} \frac{\left[\beta^2\left(2\alpha^2\rho_ct^2-1\right)+2\alpha^2(\rho_c-V_*) \right] \label{AP5}
\left(\beta^2+2V_*\alpha^2 \right)}{4\alpha^4 (1+\beta^2t^2)}.
\end{equation}
One can easily verify than equations (\ref{AP3}) and (\ref{AP5}) are equivalent only if   
\begin{equation}
\beta^2=2\alpha^2(\rho_c-V_*)=6\pi G  \rho_c (1+w')^2. 
\end{equation}
Therefore we obtain 
\begin{equation}
\phi(t)=\frac{ \text{arcsinh} \left[ \sqrt{6\pi G \rho_c}(1+w') t \right]}{\sqrt{6\pi G(1+w')}}. \label{AP6}
\end{equation}
Moreover equation (\ref{AP3}) can be written now as 
\begin{equation}
\frac{\dot{p}}{p} = \frac{4}{3}\frac{1}{(1+w')} \frac{6\pi G  \rho_c (1+w')^2  t}{1+6\pi G  \rho_c (1+w')^2 t^2}
\end{equation}
with solution 
\begin{equation}
p(t) = p_c \left( 1+6\pi G \rho_c (1+w')^2 t^2 \right)^{\frac{2}{3(1+w')}}.
\end{equation}
Now one can interpret the parameter $w'$. Namely let us calculate expression 
for the coefficient 
\begin{equation}
w = \frac{p_{\phi}}{\rho_{\phi}}= \frac{\frac{1}{2}\dot{\phi}^2-V(\phi)}{\frac{1}{2}\dot{\phi}^2+V(\phi)}.
\end{equation}
Applying potential (\ref{AP0}) together with solution (\ref{AP6}) we find 
\begin{equation}
w = w'. \label{AP7}
\end{equation}
It is now transparent that solutions found in section \ref{sec:analytical} are recovered.
Thanks to relation (\ref{AP7}) we have $dw/dt=0$, therefore all found trajectories
follow the $w=$ const.

\end{document}